\documentclass[journal,twocolumn]{IEEEtran}
\usepackage[T1]{fontenc}
\usepackage{cite}
\usepackage{amsthm,amssymb,bm}
\usepackage[tbtags]{amsmath}
\usepackage{mathrsfs}
\usepackage{graphicx}
\usepackage{enumitem}
\usepackage{algorithm}
\usepackage{algorithmic}
\usepackage{array}
\usepackage{psfrag}
\usepackage{booktabs}
\usepackage{tabularx}
\usepackage{xcolor}
\usepackage[caption=false,font=normalsize,labelfont=sf,textfont=sf]{subfig}
\usepackage{cuted}
\allowdisplaybreaks[4]
\usepackage[utf8]{inputenc}



\newtheorem{IEEEdefinition}{Definition}

\begin{document}
	\renewcommand{\qedsymbol}{}
	
	%
	%
	\title{\fontsize{20}{26}\selectfont Continuous Wavefront Design via Virtual Point Sources: A Holographic Paradigm for Near-Field XL-MIMO}
	
	%
	%
	\author{ 
	Xiyuan Liu,
	Qingqing Wu,
	Rui Wang,
	Qiaoyan Peng,
	and Jun Wu
	\thanks{X. Liu and R. Wang are with the College of Electronics and Information Engineering, Tongji University, Shanghai 201804, China, and also with the Shanghai Institute of Intelligent Science and Technology, Tongji University, Shanghai 201804, China (e-mail: 1910670@tongji.edu.cn; ruiwang@tongji.edu.cn); Q. Wu is with the Department of Electronic Engineering, Shanghai Jiao Tong University, Shanghai 200240, China (e-mail: qingqingwu@sjtu.edu.cn); Q. Peng is with the Department of Electronic Engineering, Shanghai Jiao Tong University, Shanghai 200240, China, and also with the State Key Laboratory of Internet of Things for Smart City, University of Macau, Macao SAR, China (email: qiaoyan.peng@connect.um.edu.mo); J. Wu is with the School of Computer Science, Fudan University, Shanghai 200433, China, and also with the Shanghai Qi Zhi Institute, Shanghai 200030, China (e-mail: wujun@fudan.edu.cn).}

}

	\markboth{IEEE TRANSACTIONS ON WIRELESS COMMUNICATIONS,~VOL. XX, NO.
		XX, MONTH 2025}%
	{LIU \MakeLowercase{\textit{et al.}}: A Holographic Paradigm for XL-MIMO Beamforming}

	\IEEEtitleabstractindextext{%
		\begin{abstract}
			Beamforming design for extremely large-scale multiple-input multiple-output (XL-MIMO) systems is challenging due to prohibitive computational complexity and complex near-field propagation effects.
			To address this, this paper introduces a holographic beamforming paradigm that reformulates the design from optimizing variables at spatially discrete antenna locations to shaping a continuous electromagnetic wave function over the array aperture, effectively mitigating the growth of algorithmic complexity as the array scale increases.
			We apply this paradigm to the challenging dual near-field (DNF) scenario, where strong transceiver coupling severely degrades conventional iterative algorithms.
			In this case, we propose a novel Virtual Point Source (VPS) method, which approximates the ideal wave function with a single and analytically tractable spherical-wave.
			A rigorous geometric–optical analysis is provided to show that the optimal VPS location can be determined in a fully non-iterative manner, thus decoupling the coupled DNF problem.
			The proposed method is demonstrated in an intelligent reflecting surfaces (IRS)-assisted system, where simulation results show that our non-iterative approach achieves performance comparable to converged alternating-optimization (AO) algorithms ,while incurring significantly lower complexity and avoiding convergence uncertainty
			This work offers a new theoretical framework for holographic beamforming design in XL-MIMO systems.
		\end{abstract}
		
		}

	\maketitle

	\IEEEdisplaynontitleabstractindextext
	\IEEEpeerreviewmaketitle
	
	\section{Introduction}
	The sixth generation (6G) of wireless systems is envisioned to support transformative applications demanding unprecedented data rates and connection density \cite{alsabah_ref4, alsabah_ref36}. Meeting these ambitious requirements necessitates a vast expansion of available bandwidth, which establishes high-frequency communications in the millimeter-wave (mmWave) and terahertz (THz) bands as a critical and foundational technology \cite{ref10, ref12, alsabah_ref121, alsabah_ref158}. The need to support massive machine-type connectivity for the internet of things (IoT) poses a key challenge that exceeds the capabilities of high-capacity multiple access techniques, such as frequency division multiple access (FDMA) and space division multiple access (SDMA) \cite{ref3, ref4, ref5}.
	
	However, the transition to higher frequencies introduces the fundamental challenge of severe propagation loss. To compensate for this high attenuation, substantial beamforming gains are required \cite{ref7, wang_ref4}. This requirement dictates a fundamental divergence in array architecture compared to low-frequency systems. High-frequency communication inherently relies on extremely large-scale multiple-input multiple-output (XL-MIMO) systems \cite{alsabah_ref76,  ref12}, whereas conventional low-frequency systems employ smaller arrays.
	
	This distinction in scale is driven by two primary factors. First, as antenna element spacing is proportional to wavelength, the dramatically shorter wavelengths at high frequencies enable the integration of a massive number of antennas within a feasible physical aperture \cite{alsabah_ref138, wang_ref45, wang_ref151}. Second, the channel characteristics mandate this massive scale. Low-frequency signals benefit from low attenuation and rich multipath propagation, mitigating the need for high-gain arrays. Conversely, high-frequency signals suffer from high attenuation and exhibit sparse channel properties, with power predominantly concentrated in the line-of-sight (LoS) path \cite{wang_ref71, wang_ref136}. Therefore, XL-MIMO is essential to concentrate power towards these sparse paths to counteract the severe propagation loss. This explains the stark contrast in array scale between high and low-frequency systems.
	
	The resulting large electrical aperture of high-frequency XL-MIMO systems fundamentally alters the underlying physics of wave propagation at both the transmitter and receiver. Any finite-aperture array inevitably causes diffraction. However, the impact of this diffraction changes dramatically. In conventional low-frequency systems, operating with smaller electrical apertures, diffraction effects are significant and cannot be ignored. In stark contrast, the large electrical aperture of high-frequency XL-MIMO renders diffraction effects practically negligible \cite{ref12}. This physical change, combined with the aforementioned channel sparsity, enables a complete paradigm shift in beamforming design \cite{ wang_ref71}.
	
	The aforementioned paradigm shift involves a move from an antenna-centric model to a path-centric one. In low-frequency, multipath-rich channels, the wave field at the receiver is a complex superposition of many interfering paths. Characterizing this field with a unified geometric model (e.g., a single wavefront) is ineffective, as the interfering components invalidate such a simple description. The design must therefore be antenna-centric, operating on the parameters at spatially-discrete antenna locations. Consequently, the channel is typically described through mathematical decompositions, such as its eigenvalues and eigenvectors, which lack a direct geometric interpretation. In high-frequency, sparse channels, the opposite is true. The wave field is dominated by a single path, allowing the entire array to be accurately described by a few geometric parameters, such as its direction and curvature \cite{alsabah_ref76, wang_ref138}. This enables a path-centric design, where the objective is to shape a spatially-continuous wavefront, enabling a more efficient and lightweight parameterized design \cite{wang_ref70}.
	
	The deployment of such XL-MIMO systems, however, introduces a second major challenge: near-field propagation effects become dominant. Without loss of generality, we adopt the Rayleigh distance, i.e., $d_R = 2D^2/\lambda$, where $D$ is the physical aperture (or physical size) of the array and $\lambda$ is the wavelength, to highlight the difference in near-field behavior between high and low-frequency systems, as it effectively demonstrates that the near-field region expands significantly with the large electrical aperture $D/\lambda$ \cite{wang_ref136, wang_ref151, ref11}, where $D/\lambda$ is the electrical aperture of the array. Consequently, transceivers are frequently placed in the near-field region \cite{ref10, alsabah_ref123, wang_ref45}. The planar-wave assumption applicable in the far-field becomes invalid in the near‑field due to the spherical-wavefront characteristics of electromagnetic propagation \cite{ref12, wang_ref70}.

	It is crucial to distinguish the nature of near-field effects at high versus low frequencies. In low-frequency systems, near-field effects arise mainly because the array’s physical size is not negligible compared with the communication distance. This spatial scale dictates that both the amplitude attenuation and the phase gradient vary significantly across the array, necessitating that this amplitude variation be accounted for in the beamforming design. Here, both signal amplitude and phase vary significantly across the array, and this amplitude variation must be accounted for in the beamforming design. In high-frequency systems, the near-field is fundamentally a consequence of the short wavelength ($\lambda$). As derived from the Green's function, phase variation is a function of $\lambda$, while amplitude variation is not \cite{wang_ref45, wang_ref94, wang_ref151}. This implies that high-frequency near-field beamforming is primarily a phase-only design problem. Furthermore, the term ``spherical-wave'' carries different meanings. In the low-frequency near-field, it implies a geometrically obvious large curvature. In the high-frequency near-field, the wavefront is geometrically almost planar. Its ``sphericity'' is a high-spatial-resolution phenomenon, where minute phase deviations from a perfect plane wave are small but critical and cannot be ignored. This subtle, non-geometric sphericity presents a unique challenge \cite{wang_ref71}.
	
	The near-field enables new opportunities, such as location division multiple access (LDMA) \cite{ref13} and advanced localization \cite{ ref15}. However, the coupling of angle and distance parameters in the channel model complicates algorithm design \cite{ref16, ref17, wang_ref78}. This challenge is particularly evident in systems using intelligent reflecting surfaces (IRS) \cite{ref18, ref19}. IRS have emerged as a technology that inherently integrates high-frequency operation, XL-MIMO, and near-field characteristics \cite{ref20, ref21, alsabah_ref23, wang_ref17}. As high-frequency systems are highly dependent on LoS paths, an IRS is vital for providing a virtual LoS path when the direct path is obstructed \cite{ alsabah_ref101, wang_ref18}. This deployment often leads to a dual near-field (DNF) scenario. In this case, the base station (BS) and the IRS, both being XL-MIMO systems, are in close proximity and thus in the near-field of each other \cite{ref22, wang_ref78}. While this DNF channel offers high-rank properties, the strong coupling between the transceivers makes beamforming design exceptionally complex. Consequently, conventional iterative methods like alternating optimization (AO) become sensitive to initialization, computationally intensive, and prone to converging to suboptimal solutions \cite{ref23, ref24, ref25, Williams2021_MultiuserMIMO, Williams2020_LISModel, wang_ref110}.
	
	To address these challenges, this paper develops a new beamforming design paradigm. The main contributions of this paper are summarized as follows:
	\begin{itemize}
		\item We present a fundamental analysis of the beamforming problem from a continuous-space perspective. By separating the angular spectrum into an interference spectrum and a diffraction spectrum, we analyze their independent evolution as the electrical aperture increases. This leads to the discovery of the diffraction degradation principle in the XL-MIMO limit, which provides the theoretical foundation for achieving computational complexity that is independent of the array scale.
		\item We propose and clarify the concept of the holographic multiple-input multiple-output (HMIMO). Unlike traditional definitions based on sub-wavelength dense sampling, our model is rigorously defined by two limiting conditions: an infinite electrical aperture ($D/\lambda \to \infty$) and Nyquist spatial sampling ($d=\lambda/2$). This refined definition establishes a new and physically precise framework for XL-MIMO analysis.
		\item We propose a continuous electromagnetic wave function (CWF) based holographic methodology, which leverages this holographic array model. This approach reformulates the high-dimensional, discrete optimization problem into the task of shaping a CWF over the array aperture, thereby enabling design algorithms whose complexity is independent of the number of array elements.
		\item By applying this methodology to the challenging DNF scenario, we propose the virtual point source (VPS) algorithm. The VPS method approximates the complex optimal CWF with a single, analytically tractable spherical-wave and utilizes a geometric-optical method to find the optimal source location non-iteratively, achieving a decoupled and low-complexity solution for DNF beamforming.
	\end{itemize}
	
	The remainder of this paper is organized as follows. Section II establishes an analytical framework for XL-MIMO from a continuous-space perspective. Section III details the proposed holographic methodology and introduces the VPS method. In Section IV, the proposed framework is applied to solve the DNF problem in an IRS-assisted system. Section V provides simulation results to validate our theoretical findings. Finally, Section VI concludes the paper.
	
	\textit{Notation:} Throughout this paper, scalars are denoted by italic letters (e.g., $k$), vectors by bold lowercase letters (e.g., $\bm{\omega}$), and matrices by bold uppercase letters (e.g., $\bm{H}$).
	The superscripts $(\cdot)^H$, $(\cdot)^T$ and  $(\cdot)^*$ denote the conjugate transpose, transpose, and complex conjugate, respectively. $||\cdot||_2$ represents the Euclidean norm.
	$\mathcal{F}\{\cdot\}$ denotes the Fourier Transform, and $\circledast$ denotes the convolution operation.
	$\mathbb{C}^{M \times N}$ represents the space of $M \times N$ complex matrices.

	\section{A Continuous-Space Analytical Framework for XL-MIMO}
	In this section, we establish an analytical framework for XL-MIMO from a continuous-space perspective. We first investigate the physical impacts of finite apertures, namely interference and diffraction, and subsequently introduce the foundational principle of diffraction degradation that emerges in the XL-MIMO limit.
	\subsection{Beamforming Model for Finite Apertures}
	In point-to-point multiple-input multiple-output (MIMO) systems, the received signal is conventionally described by the standard vector-matrix model, which captures the interaction between discrete antenna arrays:
	\begin{equation} \label{eq_mimo_signal_model}
		y = \bm{\omega}_r^H \bm{H} \bm{\omega}_t s + \bm{\omega}_r^H \bm{n},
	\end{equation}
	where $s$ is the transmitted symbol, $\bm{\omega}_t \in \mathbb{C}^{N_t \times 1}$ and $\bm{\omega}_r \in \mathbb{C}^{N_r \times 1}$ are the beamforming and combining vectors, respectively, $\bm{H} \in \mathbb{C}^{N_r \times N_t}$ represents the channel response matrix between the transmitter and receiver, where $N_t$ and $N_r$ are the number of transmit and receive antennas, and $\bm{n}$ is the noise vector.
	To analyze the impact of the array's finite electrical aperture from a wave-physics perspective, we first define an idealized, infinite-length beamforming sequence, $\{\omega_i[n_i]\}_{n_i=-\infty}^{\infty}$, where the subscript $i \in \{t, r\}$ pertains to either the transmitter or the receiver. This idealized construct represents a spatial excitation pattern possessing an intrinsic angular spectrum. The practical beamforming vector $\bm{\omega}_i$, which is constrained by the finite aperture, is then modeled as a truncated version of this ideal sequence. This truncation is achieved by applying a spatial window function $W_i[n_i]$, and the $n_i$-th element of the resulting vector is given by
	\begin{equation} \label{eq_vector_sequence_relation}
		[\bm{\omega}_i]_{n_i} = W_i[n_i] \omega_i[n_i],
	\end{equation}
	where $W_i[n_i]$ is non-zero only for indices corresponding to physical antenna elements.
	The introduction of the infinite sequence allows us to mathematically separate the desired beam pattern from the physical constraints of the array's finite size.
	The beamforming problem can thus be reformulated from an end-to-end channel perspective to a more physically intuitive task of shaping and matching intermediate wave fields.
	To this end, we define the incident signal wave vector at the receiver array as $\bm{\xi}_r$, which can be written as
	\begin{equation} \label{eq_xi_r_def}
		\bm{\xi}_r = \bm{H} \bm{\omega}_t.
	\end{equation}
	Similarly, leveraging the principle of channel reciprocity, a corresponding incident wave vector $\bm{\xi}_t$ is defined at the transmitter, representing the effective wave field observed from the receiver's transmission, which is
	\begin{equation} \label{eq_xi_t_def}
		\bm{\xi}_t = \bm{H}^H \bm{\omega}_r.
	\end{equation}
	The vectors $\bm{\xi}_r$ and $\bm{\xi}_t$ describe the electromagnetic fields as sampled by the receiver and transmitter arrays, respectively.
	The received power is then expressed as the inner product of the receiver's combining vector and the incident wave vector:
	\begin{equation} \label{eq_scalar_power_chap5}
		P_{rx} = \bm{\omega}_r^H \bm{\xi}_r.
	\end{equation}
	To facilitate spectral analysis, we consider the underlying infinite-length sequences $\{\omega_i[n_i]\}_{n_i \in \mathbb{Z}}$ and $\{\xi_i[n_i]\}_{n_i \in \mathbb{Z}}$, from which the finite vectors are truncated. The index $n_i$ represents the discrete spatial sample index ($i \in \{t, r\}$).  To analyze this scalar quantity in the spectral domain, we transform it into an equivalent constant sequence, $z_i[m]$, where $m \in \mathbb{Z}$ is the new sequence index. This sequence is defined to be constant for all $m$, with its value equal to $P_{rx}$. This transformation is expressed as the convolution of the term-by-term product $(\omega_i^* \xi_i)$ with an all-ones sequence, denoted as $\mathbf{1}$, as given by
	\begin{equation} \label{eq_z_sequence_construction}
		z_i[m] = ((\omega_i^* \xi_i) \circledast \mathbf{1})[m] = \sum_{n_i=-\infty}^{\infty} \omega_i^*[n_i]\xi_i[n_i], \quad i \in \{t, r\}.
	\end{equation}
	This formulation allows us to analyze the maximization of a scalar gain by examining the spectral properties of the sequence $z_i[m]$, thereby revealing the impact of the array's aperture on the beamforming problem.
	
	\subsection{Plane-Wave Interference and Diffraction}
	The angular spectrum provides a powerful framework for analyzing wave phenomena by decomposing a complex wavefront into a superposition of constituent plane waves.
	This perspective is particularly insightful for understanding how an array interacts with the electromagnetic field.
	The normalized angular spectrum, $G(\beta)$, of a discrete spatial sequence $g[n]$ with element spacing $d$ is defined as its Discrete-Time Fourier Transform (DTFT), normalized for power:
	\begin{equation} \label{eq_angular_spectrum_def}
		G(\beta) = \frac{1}{\sqrt{N}} \sum_{n=-\infty}^{\infty} g[n] e^{-\jmath k \beta n d},
	\end{equation}
	where $k=2\pi/\lambda$ is the wavenumber and $\beta = \sin\theta$ is the direction sine, with $\theta$ being the angle relative to the array normal and $N$ is the number of antenna elements. 
	The variable $\beta$ maps the physical angular domain to the spatial frequency domain, and the angular spectrum quantifies the magnitude and phase of the plane-wave component for each direction $\beta$.
	
	The angular spectrum of a transceiver array is shaped by a cascade of physical effects.
	The following derivation applies to both the transmitter and receiver; due to channel reciprocity, the principles are symmetric.
	We define this as the interference spectrum, $\Omega_i(\beta_i)$, by taking the DTFT of the sequence over the visible region $\beta_i \in [-1, 1]$:
	\begin{equation} \label{eq_interference_spectrum}
		\Omega_i(\beta_i) = \sum_{n_i=-\infty}^{\infty} \omega_i[n_i] e^{-\jmath k \beta_i n_i d_i}, \quad i \in \{t, r\}.
	\end{equation}
	Subsequently, the discrete spatial sampling of the sequence with period $d_i$ results in the periodic extension, or aliasing, of this interference spectrum.
	The resulting spectrum, $\Omega_{i, \text{aliased}}(\beta_i)$, is a periodic summation of the base interference spectrum:
	\begin{equation} \label{eq_aliased_spectrum}
		\Omega_{i, \text{aliased}}(\beta_i) = \frac{1}{d_i}\sum_{l=-\infty}^{\infty} \Omega_i\left(\beta_i - l \frac{\lambda}{d_i}\right), \quad i \in \{t, r\}.
	\end{equation}
	Finally, the finite physical aperture of the array, described by the window function $W_i$ in (\ref{eq_vector_sequence_relation}), introduces diffraction.
	In the spectral domain, this corresponds to a convolution of the aliased interference spectrum with a diffraction spectrum.
	The diffraction spectrum, $\mathcal{F}\{W_i\}(\beta_i)$, is the continuous Fourier transform of the spatial window function $W_i(x)$, where $x$ represents the continuous spatial coordinate along the array aperture. This transform is given by
	\begin{equation} \label{eq_diffraction_spectrum_relation}
		\mathcal{F}\{W_i\}(\beta_i) = \int_{-\infty}^{\infty} W_i(x) e^{-\jmath k \beta_i x} dx, \quad i \in \{t, r\}.
	\end{equation}
	For a uniform linear array (ULA) of physical length $D_i$, the window function is a rectangular pulse, and its Fourier transform yields a sinc function:
	\begin{equation} \label{eq_diffraction_spectrum_sinc}
		\mathcal{F}\{W_i\}(\beta_i) = D_i \text{sinc}\left(\frac{D_i \beta_i}{\lambda}\right), \quad i \in \{t, r\},
	\end{equation}
	where the unnormalized sinc function is defined as $\text{sinc}(z) \triangleq \sin(\pi z) / (\pi z)$.
	Thus, the complete angular spectrum generated by a physical array is the convolution $\Omega_{i, \text{aliased}}(\beta_i) \circledast \mathcal{F}\{W_i\}(\beta_i)$.
	
	The channel itself acts as a spatial filter and a mapping function from the transmitter's angular domain to the receiver's.
	This process can be conceptualized in three stages. The channel possesses spatial filters, $\Gamma_t(\beta_t)$ and $\Gamma_r(\beta_r)$, which define ``passbands'' for the wave fields.
	Only the portion of the transmitted spectrum within the passband of $\Gamma_t(\beta_t)$ propagates effectively.
	These filters can be modeled as ideal, with a value of 1 in the passband and 0 otherwise:
	\begin{equation} \label{eq_spatial_filter}
		\Gamma_i(\beta_i) = 
		\begin{cases} 
			1, & \beta_i \in \text{Passband} \\
			0, & \beta_i \in \text{Stopband} 
		\end{cases}, \quad i \in \{t, r\}.
	\end{equation}
	In addition, the channel geometry dictates a mapping relationship, $\Psi$, that transforms the filtered transmitted spectrum, $\mathcal{S}_t(\beta_t)$, to an incident spectrum at the receiver, $\mathcal{S}_r(\beta_r)$:
	\begin{equation} \label{eq_channel_mapping}
		\mathcal{S}_r(\beta_r) = \Psi(\mathcal{S}_t(\beta_t)).
	\end{equation}
	This mapped wave, upon arriving at the finite receiver array, undergoes diffraction once more due to the receiver's aperture.
	Combining these phenomena, we can formulate the complete expression for the angular spectrum of the incident wave sequence at the receiver, denoted as $\Xi_r(\beta_r) = \mathcal{F}\{\xi_r[n_r]\}$.
	It is the result of the transmitter's spectrum being aliased, convolved with the transmitter's diffraction spectrum, filtered by $\Gamma_t$, mapped by $\Psi$, and finally convolved with the receiver's diffraction spectrum:
	\begin{equation} \label{eq_full_Xi_spectrum}
		\begin{split}
			\Xi_r(\beta_r) = & \left[ \Psi \left( \left\{ \Omega_{t, \text{aliased}}(\beta_t) \circledast \mathcal{F}\{W_t\}(\beta_t) \right\} \Gamma_t(\beta_t) \right) \right] \\
			& \circledast \mathcal{F}\{W_r\}(\beta_r).
		\end{split}
	\end{equation}
	The optimal condition for reception is maximal ratio combining (MRC), which dictates that the receiver's beamforming should be matched to the incident wave, i.e., $\omega_r[n_r] \propto \xi_r^*[n_r]$, implying $\Omega_r(\beta_r) \propto \Xi_r^*(\beta_r)$.
	The  matching depends critically on whether the spatial channel is wideband or narrowband.
	
	In a spatial wideband scenario, corresponding to near-field communication, the channel filters $\Gamma_i(\beta_i)$ have a wide passband.
	For instance, the near-field channel steering vector for a point source at $(r, \theta)$ has a phase term that depends non-linearly on the antenna index, and its angular spectrum is spread over a wide range of $\beta$:
	\begin{equation}
		\bm{a}_{\text{NF}}(r, \theta) = \left[ e^{-j k (\|\bm{p}_1 - \bm{s}\| - r)}, \dots, e^{-j k (\|\bm{p}_N - \bm{s}\| - r)} \right]^T,
	\end{equation}
	where ${\bm p}_n$ is the position vector of the $n$-th antenna element and $\bm s$ is the position vector of the point source.
	In this case, the final spectrum $\Xi_r(\beta_r)$ in \eqref{eq_full_Xi_spectrum} is a non-trivial function that depends on the entire shape of the transmitted spectrum.
	Consequently, the optimal receiver beamformer $\Omega_r$ depends on the transmitter's beamformer $\Omega_t$.
	By channel reciprocity, the optimal $\Omega_t$ similarly depends on $\Omega_r$, creating a fundamental coupling in the beamforming design.
	
	Conversely, in a spatial narrowband scenario, corresponding to far-field communication, the channel filters are effectively delta functions, selecting a single spatial frequency, expressed as
	\begin{equation} \label{eq_narrowband_filter}
		\Gamma_i(\beta_i) = \delta(\beta_i - \beta_{i,0}), \quad i \in \{t, r\},
	\end{equation}
	where $\delta(\cdot)$ denotes the Dirac delta function.
	The far-field channel steering vector corresponds to a pure plane wave with a linear phase progression:
	\begin{equation}
		\bm{a}_{\text{FF}}(\beta) = \left[ 1, e^{-\jmath k d \beta}, \dots, e^{-\jmath k (N-1) d \beta} \right]^T.
	\end{equation}
	When the transmitted spectrum is filtered by $\Gamma_t(\beta_t) = \delta(\beta_t - \beta_{t,0})$, the multiplication effectively samples the transmitter's spectrum at $\beta_{t,0}$, generating an ideal plane wave.
	After propagation, this wave arrives at the receiver from a corresponding direction $\beta_{r,0}$ with a complex amplitude $C$.
	The spectrum of this incident wave is $C \delta(\beta_r - \beta_{r,0})$.
	Substituting this into the final convolution stage of (\ref{eq_full_Xi_spectrum}), the incident spectrum at the receiver becomes:
	\begin{equation} \label{eq_narrowband_incident_spectrum}
		\begin{split}
			\Xi_r(\beta_r) & = \left( C \delta(\beta_r - \beta_{r,0}) \right) \circledast \mathcal{F}\{W_r\}(\beta_r) \\
			& = C \mathcal{F}\{W_r\}(\beta_r - \beta_{r,0}).
		\end{split}
	\end{equation}
	This result shows that the incident spectrum is simply the receiver's own diffraction pattern, scaled and shifted to the direction of arrival $\beta_{r,0}$.
	This spectrum is independent of the original shape of the transmitted spectrum $\Omega_t(\beta_t)$.
	Consequently, the optimal receiver beamformer is uniquely determined by matching this incident spectrum:
	\begin{equation} \label{eq_narrowband_mrc}
		\Omega_r(\beta_r) \propto \Xi_r^*(\beta_r) \propto \left( \mathcal{F}\{W_r\}(\beta_r - \beta_{r,0}) \right)^*.
	\end{equation}
	In this scenario, the optimal beamforming designs at the transmitter and receiver are decoupled; each only needs to align with its respective channel direction.
	
	\subsection{Spherical-Wave Channel Model}
	An alternative perspective for near-field communications is the decomposition of wave fields into spherical-waves (or Green's functions).
	In this domain, the spatial frequency variable is a spatial point $\bm{s}$ representing the origin of a spherical-wave, and the channel's spatial filters, $\Gamma_i$, are functions of this coordinate.
	Let $\Omega_i(\bm{s})$ be the beamforming response function in this spherical-wave domain. The channel's spatial filter, $\Gamma_i(\bm{s})$, is non-zero only within a specific ``Focal Region''. This term represents the spatial volume containing all potential spherical-wave origins, denoted by $\bm{s}$, that serve as effective foci for the array. This filter can be idealized as
	\begin{equation}
		\Gamma_i(\bm{s}) = 
		\begin{cases} 
			1, & \bm{s} \in \text{Focal Region} \\
			0, & \text{otherwise} 
		\end{cases}, \quad i \in \{t, r\}.
	\end{equation}
	
\begin{figure}[!t]
		\centering
		\includegraphics[width=0.75\columnwidth]{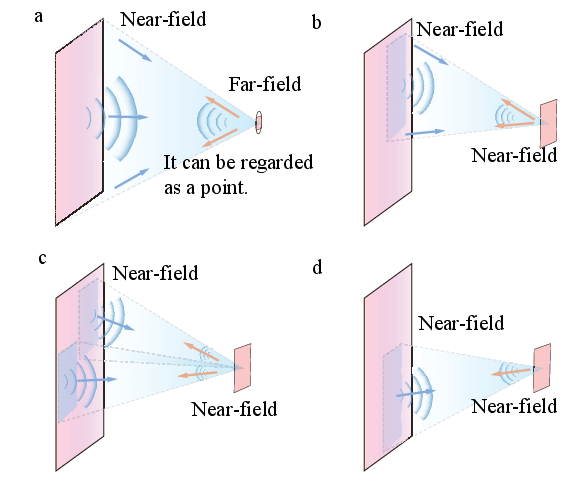}
		\caption{Illustration of (a) SNF and (b)-(d) DNF scenarios.}
		\label{fig_dual_near_chap5}
	\end{figure}

	The distinction between narrowband and wideband channels also translates to this perspective.
	A spatial narrowband channel in this domain corresponds to a filter that is non-zero only at a single point, i.e., $\Gamma_i(\bm{s}) = \delta(\bm{s} - \bm{s}_0)$.
	This scenario is characteristic of single near-field (SNF) communication, where one device is a point-like object in the near-field of a large array (see Fig. \ref{fig_dual_near_chap5}(a)).
	The ideal incident wave spectrum is a delta function centered at $\bm{s}_0$, but this wave undergoes diffraction due to the finite array aperture.
	This effect is captured by a spatial diffraction function, $\mathcal{D}_i(\bm{s})$.
	The actual incident wave spectrum at the array, $\Xi_i(\bm{s})$, is the convolution of the ideal spectrum with this diffraction function.
	Therefore, the optimal beamforming response function must match this resulting spectrum:
	\begin{equation} \label{eq_snf_beamforming}
		\Omega_i(\bm{s}) \propto \Xi_i^*(\bm{s}) \propto \left(\delta(\bm{s} - \bm{s}_0) \circledast \mathcal{D}_i(\bm{s})\right)^*.
	\end{equation}
	The design remains decoupled as it only depends on the fixed location $\bm{s}_0$ and the array's own diffraction properties.
	A spatial wideband channel, in contrast, corresponds to a filter whose focal region is a continuous volume.
	This is the case in DNF communication, where two large-aperture arrays are in each other's near-field (see Fig. \ref{fig_dual_near_chap5}(b)-(d)).
	In this scenario, propagation involves a complex superposition of many spherical-waves.
	Consequently, the optimal beamforming at one array depends on the entire excitation pattern of the other, resulting in a coupled design problem.
	
	\subsection{Diffraction Degradation in XL-MIMO}
	The progression towards higher communication frequencies is driven by the demand for vast bandwidth.
	A primary challenge at these frequencies is severe path loss, which necessitates the use of XL-MIMO arrays to generate highly focused beams.
	A key consequence of moving to higher frequencies (shorter wavelength $\lambda$) is that for a fixed physical aperture $D$, the electrical aperture, $D/\lambda$, increases significantly.
	This expansion fundamentally alters the behavior of wave propagation.
	
\begin{figure}[!t]
		\centering
		\includegraphics[width=0.7\columnwidth]{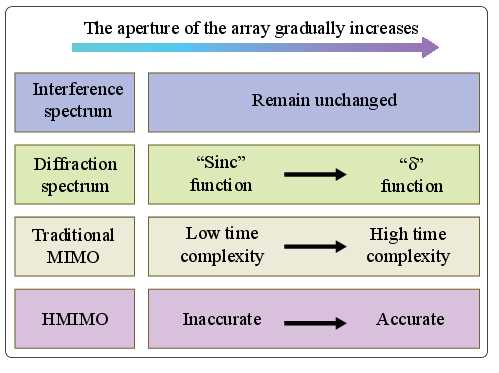}
		\caption{Principle of diffraction degradation on XL-MIMO and applicability of holographic algorithms.}
		\label{fig_trendofHMIMO_chap5}
	\end{figure}

	As the electrical aperture $D/\lambda$ becomes very large, the main lobe of the diffraction spectrum's sinc function in (\ref{eq_diffraction_spectrum_sinc}) becomes extremely narrow.
	This behavior in the limit $D/\lambda \to \infty$ is captured by the relationship
	\begin{equation} \label{eq_diffraction_degradation}
		\lim_{D/\lambda \to \infty} D_i \text{sinc}\left(\frac{D_i \beta_i}{\lambda}\right) \propto \delta(\beta_i), \quad i \in \{t, r\}.
	\end{equation}
	This phenomenon is termed diffraction degradation. Its implication is that for sufficiently large arrays, the distorting effect of the finite aperture vanishes.
	The convolution with the diffraction spectrum in (\ref{eq_full_Xi_spectrum}) becomes an identity operation.
	Consequently, the observed angular spectrum at an array converges to its underlying aliased interference spectrum.
	The spectrum of the incident wave at the receiver in (\ref{eq_full_Xi_spectrum}) simplifies to:
	\begin{equation} \label{eq_Xi_spectrum_degraded}
		\Xi_r(\beta_r) \xrightarrow{D/\lambda \to \infty} \Psi \left( \Omega_{t, \text{aliased}}(\beta_t) \Gamma_t(\beta_t) \right).
	\end{equation}
	In this limit, the problem of optimal beamforming simplifies to matching the beamformers directly to the channel's intrinsic spatial filter response, without the confounding effects of diffraction.
	This simplification decouples the algorithm's complexity from the array's physical scale, forming the theoretical basis for efficient holographic beamforming, as illustrated in Fig. \ref{fig_trendofHMIMO_chap5}.
	
	Finally, according to Weyl's identity \cite{9765526}, any spherical-wave can be represented as a superposition of plane waves:
	\begin{equation} \label{eq_weyl_identity}
		\frac{e^{jkR}}{R} = \frac{jk}{2\pi} \iint \frac{1}{k_z} e^{j(k_x x + k_y y + k_z |z|)} dk_x dk_y.
	\end{equation}
	where $R = \sqrt{x^2+y^2+z^2}$ with $(x, y, z)$ being the Cartesian coordinates. Variable $k_z$ satisfies  $k_z = \sqrt{k^2 - k_x^2 - k_y^2}$, where $k_x$ and $k_y$ are the wave vector components along the $x$ and $y$ axes.
	This identity guarantees that the principle of diffraction degradation, derived here from a plane-wave perspective, is a fundamental physical phenomenon that holds true regardless of the chosen mathematical basis.
	As diffraction effects diminish, the array gains the ability to generate arbitrarily precise wavefronts, approaching a state of perfect spatial resolution.
	In this ``holographic'' limit, the channel can be perfectly manipulated using either plane-wave or spherical-wave representations.
	
	\section{New Definitions for Holographic Arrays and the Holographic Methodology}
	
	Building on the analytical framework from the preceding section, in this section, we first provide a new and rigorous definition for the holographic array based on its limiting electrical properties. We then formally introduce the holographic methodology, a path-centric design approach that reformulates the high-dimensional discrete optimization problem into a tractable continuous-wave-function-shaping task.
	\subsection{The Concept of a Holographic Array}
	To achieve a lossless and non-redundant transformation between the spatial domain and the angular spectrum domain, the array's spatial sampling rate must adhere to the Nyquist criterion.
	Spatial undersampling, where the element spacing $d > \lambda/2$, causes aliasing in the angular spectrum, as spectral replicas overlap within the visible region $\beta \in [-1, 1]$. This aliasing phenomenon is described by (\ref{eq_aliased_spectrum})
	Aliasing makes distinct physical directions indistinguishable, resulting in a loss of design freedom in the angular domain.
	Conversely, spatial oversampling ($d < \lambda/2$) avoids aliasing but introduces coupling in the spatial domain. The phase difference between adjacent elements for an incident plane wave becomes restricted to a subset of the full $[-\pi, \pi]$ range, limiting the array's ability to generate arbitrary excitation sequences.
	Critical sampling ($d = \lambda/2$) is the condition that avoids both aliasing and spatial-domain coupling, thereby simultaneously maximizing the design freedom in both domains.
	Based on the preceding analysis, the holographic array is defined as follows.
	\begin{IEEEdefinition}[Holographic Array]
		A holographic array is an idealized model characterized by two limiting conditions: 1) its electrical aperture is infinite ($D/\lambda \to \infty$), causing diffraction effects to vanish;
	 2) its element spacing adheres to the Nyquist spatial sampling criterion ($d=\lambda/2$)~\cite{Pizzo2022_NyquistSampling}.
		The term ``holographic'' signifies the array's theoretical ability to perfectly record and reconstruct the complete spatial electromagnetic wave field, enabling infinite spatial resolution.
	\end{IEEEdefinition}
	
	\begin{table}[!t]
		\renewcommand{\arraystretch}{1.3}
		\caption{Comparison of Traditional and Proposed Holographic Array Definitions}
		\label{tab:holography_comparison}
		\centering
		\begin{tabularx}{\columnwidth}{>{\raggedright\arraybackslash}p{2.5cm} >{\centering\arraybackslash}X >{\centering\arraybackslash}X}
			\toprule
			\textbf{Feature} & \textbf{Traditional Holography} & \textbf{Proposed Holography} \\
			\midrule
			\textbf{Fundamental Limit}   & Infinitely Dense Spacing ($\lambda$ is fixed) & Infinite Frequency ($D$ is fixed) \\
			\textbf{Physical Aperture $D$} & Finite & Finite \\
			\textbf{Electrical Aperture $D/\lambda$} & Finite & Infinite ($D/\lambda \to \infty$ as $\lambda \to 0$) \\
			\textbf{Element Spacing $d$} & Sub-wavelength ($d < \lambda/2$) & Nyquist Sampling ($d = \lambda/2$) \\
			\textbf{Macroscopic View} & Continuous ($d \to 0$) & Continuous ($d \to 0$) \\
			\textbf{Reason for $d \to 0$}  & By design (dense packing) & Consequence of $\lambda \to 0$ while $d=\lambda/2$ \\
			\textbf{Why 
				Holographic} & Perfect wave control via spatial continuity & Infinite spatial resolution \\
			\bottomrule
		\end{tabularx}
	\end{table}
	
	A key characteristic of the holographic array is its dual nature of being both discrete and continuous.
	At the microscopic level, it must be discrete with an element spacing of exactly a half-wavelength to ensure a bijective mapping between the spatial and wave-vector domains.
	In Table 1, we provide a detailed comparison between the conventional MIMO paradigm and our new holographic framework, highlighting the differences in their underlying physical assumptions and resulting properties.
	As detailed in Table \ref{tab:holography_comparison}, this continuity arises because its electrical aperture tends to infinity, which we argue is achieved by letting the wavelength tend to zero ($\lambda \to 0$).
	To maintain the critical sampling condition $d = \lambda/2$, the physical spacing $d$ must also approach zero, making the discrete elements so dense that the array appears continuous.
	
	The holographic property is determined by the array's electrical aperture, not its physical one.
	An array becomes holographic as $D/\lambda \to \infty$. This can be achieved by either increasing $D$ or decreasing $\lambda$.
	We focus on the latter, as decreasing $\lambda$ while keeping $D$ fixed maintains a constant physical geometry, isolating the effect of the increasing electrical aperture.
	This aligns with the research trend towards higher frequency bands.
	
	The beamforming design complexity for holographic arrays presents a stark contrast when analyzed from the antenna-centric versus the path-centric perspective. The antenna-centric methodology, with a complexity that scales directly with the number of elements $N$, becomes computationally intractable in the holographic limit as $N$ approaches infinity. Conversely, the path-centric, or wavefront, perspective becomes highly efficient. In this continuous-aperture limit, diffraction effects are negligible, allowing the array's phase configuration to be described by a rigorous geometric model. This model is parameterized by the properties of a few dominant propagation paths, such as wavefront curvature and direction, rather than by the states of the $N$ individual elements. This low-parameter, path-centric model is thus inherently tractable and forms the basis for the holographic methodology.
	
	\subsection{Definition of the Holographic Methodology}
	\begin{IEEEdefinition}[Holographic Methodology]
		A holographic methodology is a two-step, path-centric design approach that resolves the complexity challenge of continuous-aperture arrays.
		First, it leverages the diffraction-free properties of the holographic limit to characterize the continuous aperture using a simple geometric model. This step identifies the geometric parameters (e.g., wavefront curvature and direction) of the few dominant propagation paths.
		Second, this low-parameter geometric model is used as a foundation to compute and configure the specific phase states of the $N$ individual elements across the entire array.
		This methodology thereby bypasses the intractable $N$-dimensional optimization problem, reducing it to a tractable parameter estimation problem governed by the small number of dominant paths, which is fundamentally more scalable than the element-centric approach.
	\end{IEEEdefinition}
	
	The holographic methodology shifts the design perspective from optimizing individual elements to shaping the overall wave field.
	Its complexity is definied by the desired wave function, not the array's scale, making it highly efficient for XL-MIMO.
	Since the model neglects diffraction, its accuracy depends on the diffraction degradation principle, holding well for large electrical apertures.
	
\begin{figure}[!t]
		\centering
		\includegraphics[width=0.7\columnwidth]{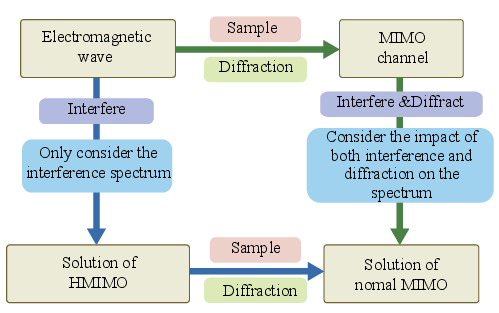}
		\caption{Difference in technical routes for beamforming design on HMIMO versus traditional MIMO.}
		\label{fig_route_chap5}
	\end{figure}

	To design the continuous wave function, we draw inspiration from the Huygens-Fresnel principle \cite{wang_ref151}, which models a complex wave field as a superposition of secondary sources.
	This can be expressed by modeling the field $U(\bm{p})$ as an integral of Green's functions \cite{9765526} over a distribution of virtual sources $\rho(\bm{s})$ within a volume $\mathcal{V}$:
	\begin{equation} \label{eq_huygens_superposition}
		U(\bm{p}) = \int_{\mathcal{V}} G(\bm{p}, \bm{s}) \rho(\bm{s}) d\bm{s},
	\end{equation}
	where $G(\bm{p}, \bm{s})$ is the Green's function for a spherical-wave from a source at $\bm{s}$.
	This suggests modeling the field as originating from an equivalent set of VPS, which provides a more compact representation of the channel's propagation modes.
	In the holographic limit, where diffraction is absent, wave propagation can be modeled by geometric optics, allowing us to use simple geometric constructions to find the optimal locations for these virtual sources.
	The fundamental difference between this methodology and conventional approaches is illustrated in Fig. \ref{fig_route_chap5}.
	Traditional methods operate directly on the discrete parameters of the physical array, complicated by effects like diffraction.
	In contrast, the holographic methodology first solves an idealized problem in a continuous domain and then maps this solution back to the physical array, effectively isolating the core wave-shaping task from the physical limitations of the array.
	
	\section{Holographic Solution for the DNF Problem}
	In this section, we apply the proposed holographic methodology to address the challenging DNF coupling problem. To tackle this, we first introduce the VPS approximation, which leads to a non-iterative, geometric-optical solution. Finally, the practical application and effectiveness of this method are demonstrated in an IRS-assisted communication system.
	\subsection{Proposed Holographic DNF Beamforming}
	The beamforming design paradigm varies significantly with the field region, which is dictated by the nature of the channel's spatial filter, $\Gamma_i$.
	In the far-field, the channel acts as a narrowband filter in the angular domain, $\Gamma_i(\beta) = \delta(\beta - \beta_{i,0})$, and the problem is decoupled.
	In the SNF scenario, the channel is narrowband from a spherical-wave perspective, with $\Gamma_i(\bm{s}) = \delta(\bm{s} - \bm{s}_0)$, and the problem is also decoupled.
	However, in the DNF, the channel is inherently wideband from both perspectives. The spatial filter $\Gamma_i(\beta)$ is a wide, continuous function, and similarly, $\Gamma_i(\bm{s})$ is non-zero over a continuous volume.
	This wideband nature of the DNF channel is the fundamental cause of beamforming coupling.
	
	\subsubsection{Mathematical Representation of DNF Coupling}
	In the conventional space-domain formulation, the beamforming design objective is to maximize the received power, $|\bm{\omega}_r^H \bm{H} \bm{\omega}_t|^2$. A standard method for this problem is the AO algorithm, which iteratively solves two coupled subproblems.
	First, for a fixed transmitter beamformer $\bm{\omega}_t$, the optimal receiver beamformer $\bm{\omega}_r$ is found by applying the MRC principle to the incident wave $\bm{H}\bm{\omega}_t$. The resulting combining vector is
	\begin{equation}\label{eq_rx_bf_opt_ao_chap4}
		\bm{\omega}_r^{\text{opt}} = \frac{\bm{H}\bm{\omega}_t}{||\bm{H}\bm{\omega}_t||_2}.
	\end{equation}
	Second, for a fixed receiver beamformer $\bm{\omega}_r$, the problem is symmetric. The optimal transmitter beamformer $\bm{\omega}_t$ is derived by applying the principle of maximal ratio transmission (MRT) to the reciprocal effective channel $\bm{H}^H \bm{\omega}_r$, yielding
	\begin{equation}\label{eq_tx_bf_opt_ao_chap4}
		\bm{\omega}_t^{\text{opt}} = \frac{\bm{H}^H\bm{\omega}_r}{||\bm{H}^H\bm{\omega}_r||_2}.
	\end{equation}
	The mutual dependency in (\ref{eq_rx_bf_opt_ao_chap4}) and (\ref{eq_tx_bf_opt_ao_chap4}), where the optimal $\bm{\omega}_r$ depends on $\bm{\omega}_t$ and vice versa, defines the coupling. This circular dependency renders the iterative AO process highly sensitive to initialization and prone to convergence at suboptimal local optima. These drawbacks are not merely algorithmic artifacts but rather highlight the fundamental intractability of the conventional antenna-centric formulation, which operates directly on a high-dimensional vector of discrete, tightly coupled variables. This challenge motivates a paradigm shift from the antenna-centric approach to the path-centric, continuous-wavefront-based methodology proposed in this paper.
	
	\subsubsection{Single VPS Approximation Rationale}
	The core idea of our proposed method is to circumvent the problem of finding the complex, unknown wave function that maximizes power
	transfer.
	Instead, we construct a simple, analytical approximation, $U_{\text{approx}}$, in the form of a single spherical-wave emanating from an optimally placed VPS.
	This effectively treats the wideband DNF channel as an approximated narrowband one.
	This approximation is justified in high-frequency, LoS-dominant near-field scenarios.
	First, the dominant characteristic of near-field propagation is the rapid spatial variation of phase, whereas amplitude variation across the array is less significant.
	This disparity is quantifiable. For a representative 28 GHz ($\lambda \approx 10.71 \text{ mm}$) system with a 1-meter physical aperture ($D=1\text{m}$), the Rayleigh distance is $d_R = 2D^2/\lambda \approx 186.7 \text{ m}$. At a typical cellular distance of $r=150 \text{ m}$, the system operates within the near-field ($r < d_R$), and the maximum phase deviation from the planar wave assumption is $\Delta\phi_{\text{max}} \approx 0.156\pi$, which significantly exceeds the $\pi/8$ far-field threshold. Conversely, the signal amplitude $A(x)$ at an element is inversely proportional to its line-of-sight propagation distance, $\mathcal{R}(x) = \sqrt{r^2 + x^2}$, where $x$ is the position relative to the array center. The resulting spatial amplitude variation is negligible; the amplitude at the array center, $A(0)$, is proportional to $1/r$, while the amplitude at the edge, $A(D/2)$, is proportional to $1/\mathcal{R}(D/2)$. The variation ratio, defined as $1 - A(D/2)/A(0)$, evaluates to $1 - (r / \mathcal{R}(D/2))$, which is only $0.00056\%$.
	This demonstrates that matching the phase of a single spherical-wave is sufficient to capture the most critical feature of the near-field channel.
	Second, a single VPS provides a simple and extensible model.
	For the LoS-dominant scenarios considered, a single spherical-wave serves as a highly effective target, closely approximating the true optimal wave function.
	This methodology can be naturally extended to handle more complex multipath environments by modeling each significant path with its own VPS, transforming a complex problem into a set of simpler geometric ones.
	
	\subsubsection{Geometric VPS Selection}
\begin{figure}[!t]
		\centering
		\includegraphics[width=0.8\columnwidth]{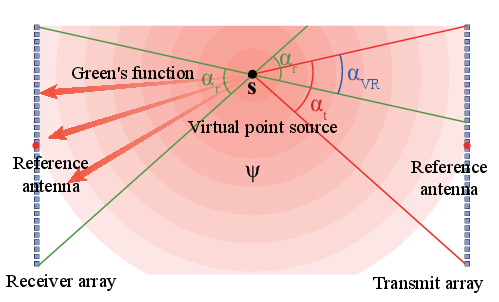}
		\caption{Decoupling beamforming performance analysis by constructing a virtual point source. The geometric gain is determined by the Common Visible Region (CVR), corresponding to the angle $\alpha_{VR}$, which is the intersection of the Visible Regions (VRs) subtended by the angles $\alpha_t$ and $\alpha_r$, respectively.}
		\label{fig_vps_decouple_chap4}
	\end{figure}

	The problem is now reduced to finding the optimal location, $\bm{s}_{\text{opt}}$, for the single VPS.
	In the holographic limit where diffraction is negligible, wave propagation is described by geometric optics.
	The goal is to select a point $\bm{s}$ that maximizes the geometric coupling between the transmitting aperture $A_t$ and the receiving aperture $A_r$.
	As illustrated in Fig. \ref{fig_vps_decouple_chap4}, for any candidate point $\bm{s}$, the transmitter and receiver arrays subtend specific solid angles, defining their individual Visible Regions (VRs).
	Let $\alpha_t(\bm{s})$ and $\alpha_r(\bm{s})$ be the angles corresponding to these VRs.
	The shared region for energy transfer is the intersection of these VRs, termed the Common Visible Region (CVR), with angle $\alpha_{VR}(\bm{s})$.
	The optimal VPS location, $\bm{s}_{\text{opt}}$, is the point that maximizes the overlap between the two VRs.
	We define a geometric coupling factor, $\eta_G(\bm{s})$, as:
	\begin{equation}\label{eq_geometric_coupling_factor}
		\eta_G(\bm{s}) = \frac{\alpha_{VR}^2(\bm{s})}{\alpha_t(\bm{s}) \alpha_r(\bm{s})}.
	\end{equation}
	The optimization problem is thus transformed into:
	\begin{equation}\label{eq_s_opt_geometric}
		\bm{s}_{\text{opt}} = \arg\max_{\bm{s}} \eta_G(\bm{s}).
	\end{equation}
	Geometric analysis shows that $\eta_G(\bm{s})$ is maximized when the VRs are perfectly aligned, i.e., $\alpha_t(\bm{s}) = \alpha_r(\bm{s}) = \alpha_{VR}(\bm{s})$.
	For two linear or planar arrays, this condition is uniquely met at the intersection of the lines connecting their opposite endpoints.
	This construction, which we term the ``opposing triangles'' method, provides a non-iterative, analytical solution for $\bm{s}_{\text{opt}}$.
	
	\subsubsection{Decoupled Performance Analysis}
	The VPS method also allows for the decoupling of factors that determine beamforming performance.
	The total received power can be analyzed as the product of three independent factors.
	The first is the Geometric Coupling Gain ($G_{\text{geom}}$), which depends on how well the chosen VPS location $\bm{s}$ aligns the two arrays and is determined by the solid angles subtended by the arrays as seen from the VPS.
	\begin{equation}
		G_{\text{geom}}(\bm{s}) \propto \text{Area}(\Omega_t(\bm{s}) \cap \Omega_r(\bm{s})).
	\end{equation}
	The second is the Focusing Gain ($G_{\text{focus}}$), which measures how effectively the beamformers are matched to the spherical-wave defined by $\bm{s}$.
	\begin{equation}
		G_{\text{focus}} = \left|
		\int \Omega_{t,r}^*(\beta) \Xi_{t,r}(\beta, \bm{s}) d\beta \right|^2,
	\end{equation}
	where $\Xi_{t,r}(\beta, \bm{s})$ is the angular spectrum of the spherical-wave from/to $\bm{s}$.
	Finally, the analysis must account for Diffraction Loss ($L_{\text{diff}}$), which represents the energy lost due to diffraction at the finite physical apertures.
	\begin{equation}
		L_{\text{diff}} \propto \left( \frac{\lambda}{D_t} \right)^2 \left( \frac{\lambda}{D_r} \right)^2.
	\end{equation}
	This decomposition provides clear physical insights and allows for the separate analysis of each performance-limiting factor.
	
	\subsection{Application to IRS-Aided Communications}
	This subsection applies the VPS method to an IRS-assisted communication scenario, where DNF conditions often exist between the BS and a large IRS.
	Consider a system where the BS and the IRS are both ULAs in a DNF setup. The received signal is:
	\begin{equation} \label{eq_irs_received_signal_chap4}
		y={\bm h}^{\rm H}_r \bm{\Theta} {\bm H}_t {\bm \omega}_t s + n,
	\end{equation}
	where $\bm{H}_t$ is the BS-IRS channel, $\bm{h}_r$ is the IRS-user channel, $\bm{\omega}_t$ is the BS beamformer, and $\bm{\Theta}$ is the IRS phase-shift matrix where $M$ is the number of IRS elements.
	The objective is to solve the joint optimization problem:
	\begin{equation} \label{eq_irs_opt_problem_full_chap4}
		\begin{split}
			\max_{\bm{\omega}_t, \bm{\Theta}} \quad & \left| {\bm h}^{\rm H}_r \bm{\Theta} {\bm H}_t {\bm \omega}_t \right|^2 \\
			\text{s.t.} \quad & ||\bm{\omega}_t||_2^2 = 1, \\
			& |[\bm{\Theta}]_{mm}| = 1, \; \forall m = 1, \dots, M.
		\end{split}
	\end{equation}
	This problem is coupled and difficult to solve directly. The VPS method decouples it by reformulating the objective.
	We introduce a VPS at location $\bm s$ between the BS and the IRS to define an ideal spherical-wave.
	The design is split into two steps. First, the BS beamformer $\bm{\omega}_t$ is designed to focus energy onto the VPS, and the IRS phases are configured to coherently combine the energy arriving from the VPS. Second, the IRS phases must also steer the reflected beam towards the user.
	
\begin{figure}[!t]
		\centering
		\includegraphics[width=0.7\columnwidth]{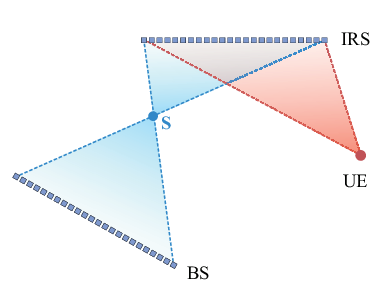}
		\caption{Constructing the virtual point source using the opposing triangles method.}     \label{fig_antitri_chap4}
	\end{figure}

	The optimal location for $\bm s$ is found by maximizing the
	BS-IRS link gain, for which we use the ``opposing triangles''
	geometric construction, as shown in Fig. \ref{fig_antitri_chap4}.
	This figure provides a visual representation of this decoupled beamforming strategy: the blue-shaded cone illustrates the ideal spherical-wavefront converging from the BS array to the VPS ($\bm s$), while the red-shaded cone illustrates the subsequent reflected wavefront, which the IRS phase-shifts to focus the energy from $\bm s$ onto the user.

	Let $\bm{p}_n$, $\bm{q}_m$, and $\bm{r}$ denote the positions of the BS antennas, IRS elements and user, respectively. The BS beamforming phase $\phi_{\rm BS}(n)$ is designed to create a wave converging at $\bm s$:
	\begin{equation}\label{VPS_BS_chap4}
		\phi_{\rm BS}(n) = k \Vert {\bm p}_n-{\bm s} \Vert.
	\end{equation}
	The phase shift of the $m$-th IRS element, $\phi_{\rm IRS}(m)$, must compensate for the phase of the incoming wave from $\bm s$ and add the phase required to focus the outgoing wave to the user at $\bm r$.
	The total required phase is the sum of the phase delays of the two paths, which effectively aligns the signals coherently at the user:
	\begin{equation}\label{VPS_IRS_chap4}
		\phi_{\rm IRS}(m) = k (\Vert {\bm s}-{\bm q}_m \Vert + \Vert {\bm q }_m - {\bm r}  \Vert).
	\end{equation}
	The non-iterative  VPS algorithm to solve problem (\ref{eq_irs_opt_problem_full_chap4}) is detailed in Algorithm \ref{alg:VPS_BF_chap4}.
	\begin{algorithm}
		\caption{VPS Beamforming Algorithm}
		\begin{algorithmic}[1]
			\STATE \textbf{Input}: BS antenna positions $\{\bm p_n\}_{n=1}^N$, IRS element positions $\{\bm q_m\}_{m=1}^M$, user position $\bm r$, wavenumber $k$.
			\STATE \textbf{Output}: BS beamforming vector $\bm \omega_t$, IRS reflection matrix $\bm \Theta$.
			\STATE \textbf{Step 1: Determine virtual point source location}
			\STATE \quad Construct opposing triangles with the BS and IRS arrays as their opposite sides.
			\STATE \quad Determine the vertex of the opposing triangles as the position of the virtual point source $\bm s$.
			\STATE \textbf{Step 2: Calculate BS beamforming phase}
			\STATE \quad \textbf{for} $n=1$ to $N$ \textbf{do}
			\STATE \quad\quad Calculate the distance from BS antenna $n$ to the virtual point source $\bm s$: $d_{\text{BS},n} = \Vert \bm p_n - \bm s \Vert$.
			\STATE \quad\quad Calculate the phase shift for BS antenna $n$: $\phi_{\rm BS}(n) = k d_{\text{BS},n}$.
			\STATE \quad \textbf{end for}
			\STATE \quad Construct the BS beamforming vector $\bm \omega_t$, whose $n$-th element is $[\bm \omega_t]_n = e^{\jmath \phi_{\rm BS}(n)}$.
			\STATE \textbf{Step 3: Calculate IRS reflection phase shifts}
			\STATE \quad \textbf{for} $m=1$ to $M$ \textbf{do}
			\STATE \quad\quad Calculate the distance from $\bm s$ to IRS element $m$: $d_{\text{IRS},m}^{\text{in}} = \Vert \bm s - \bm q_m \Vert$.
			\STATE \quad\quad Calculate the distance from IRS element $m$ to the user $\bm r$: $d_{\text{IRS},m}^{\text{out}} = \Vert \bm q_m - \bm r \Vert$.
			\STATE \quad\quad Calculate the phase shift for IRS element $m$: $\phi_{\rm IRS}(m) = k (d_{\text{IRS},m}^{\text{in}} + d_{\text{IRS},m}^{\text{out}})$.
			\STATE \quad \textbf{end for}
			\STATE \quad Construct the IRS reflection matrix $\bm \Theta$, whose $m$-th diagonal element is $[\bm \Theta]_{mm} = e^{\jmath \phi_{\rm IRS}(m)}$.
		\end{algorithmic}
		\label{alg:VPS_BF_chap4}
	\end{algorithm}
	
	\section{Simulation Results and Analysis}
	
\begin{figure}[!t]
		\centering
		\includegraphics[width=0.85\columnwidth]{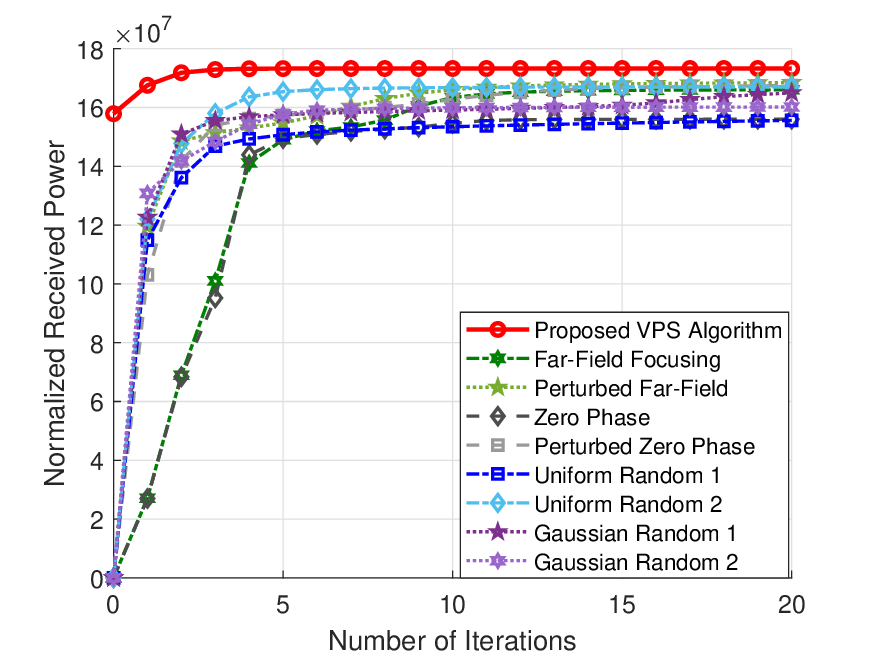}
		\caption{Convergence performance comparison of the AO algorithm under different initialization schemes.}
		\label{fig_ite_HMIMO_chap5}
	\end{figure}

\begin{figure}[!t]
		\centering
		\includegraphics[width=0.85\columnwidth]{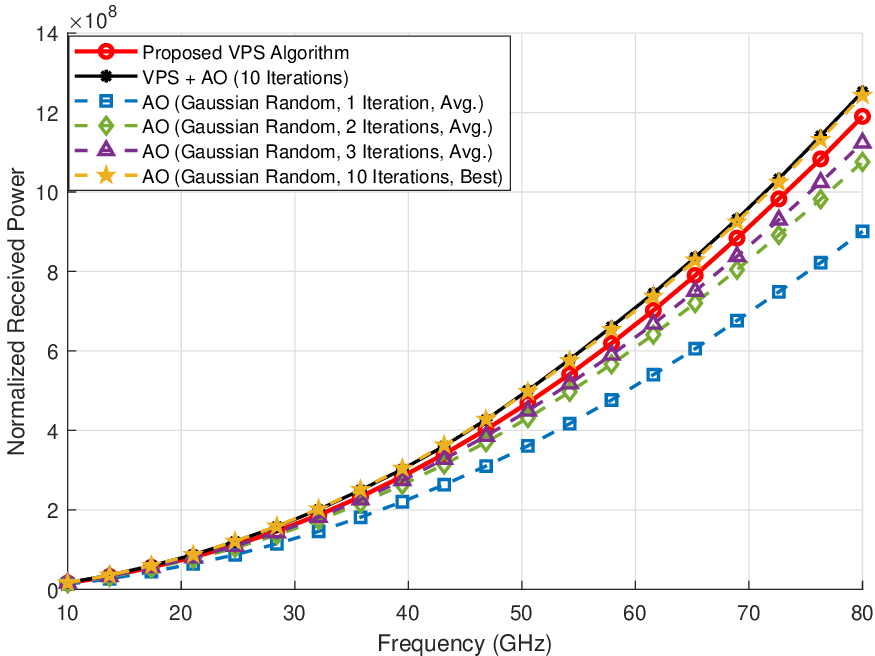}
		\caption{Normalized received power versus frequency with fixed physical size.}
		\label{fig_exp1_freq_vs_power_chap5}
	\end{figure}

\begin{figure}[!t]
		\centering
		\includegraphics[width=0.85\columnwidth]{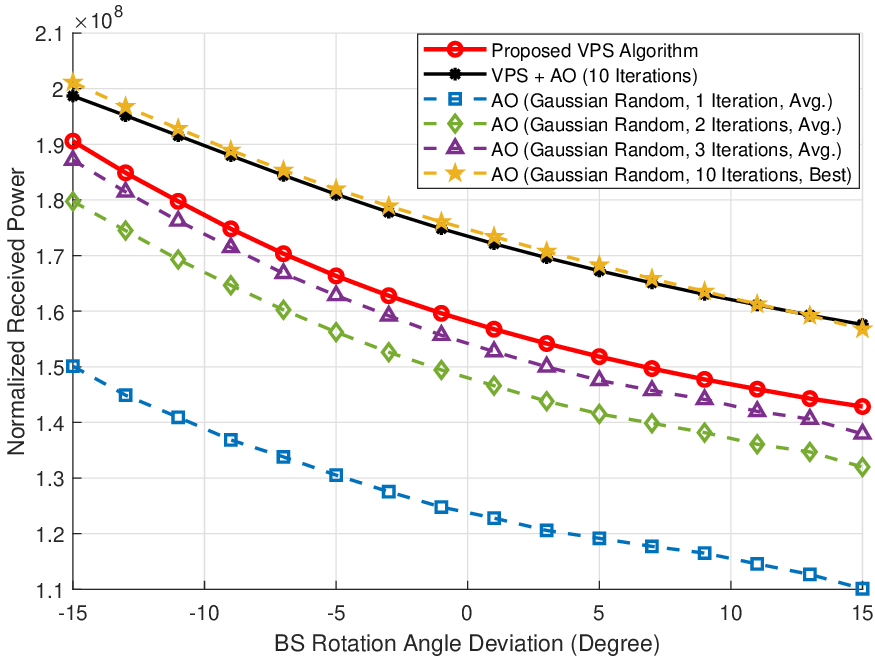}
		\caption{Normalized received power versus the rotation angle offset of the BS.}
		\label{fig_exp2_bs_angle_vs_power_chap5}
	\end{figure}

\begin{figure}[!t]
		\centering
		\includegraphics[width=0.85\columnwidth]{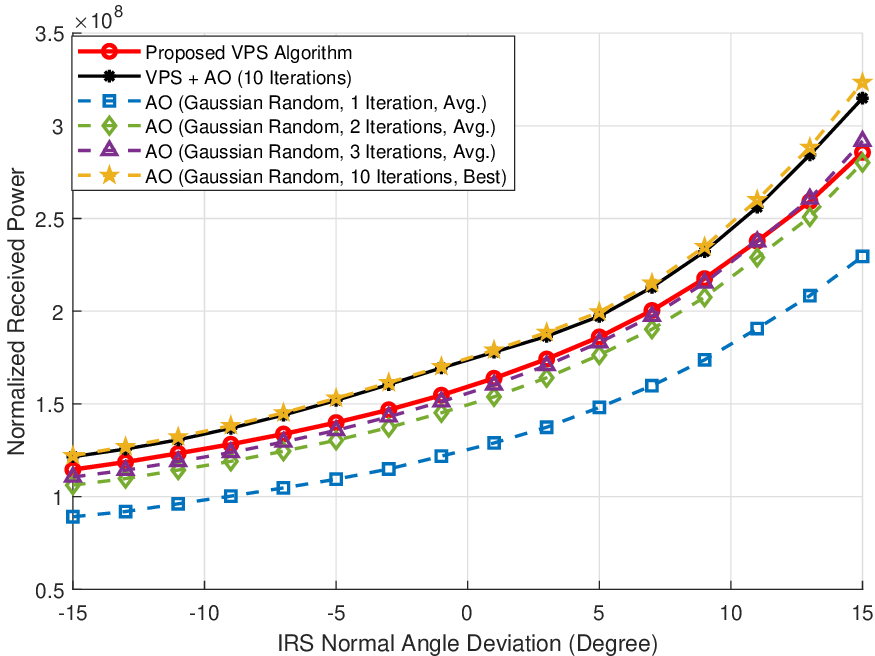}
		\caption{Normalized received power versus the normal angle offset of the IRS.}
		\label{fig_exp3_irs_angle_vs_power_chap5}
	\end{figure}

\begin{figure}[!t]
		\centering
		\includegraphics[width=0.85\columnwidth]{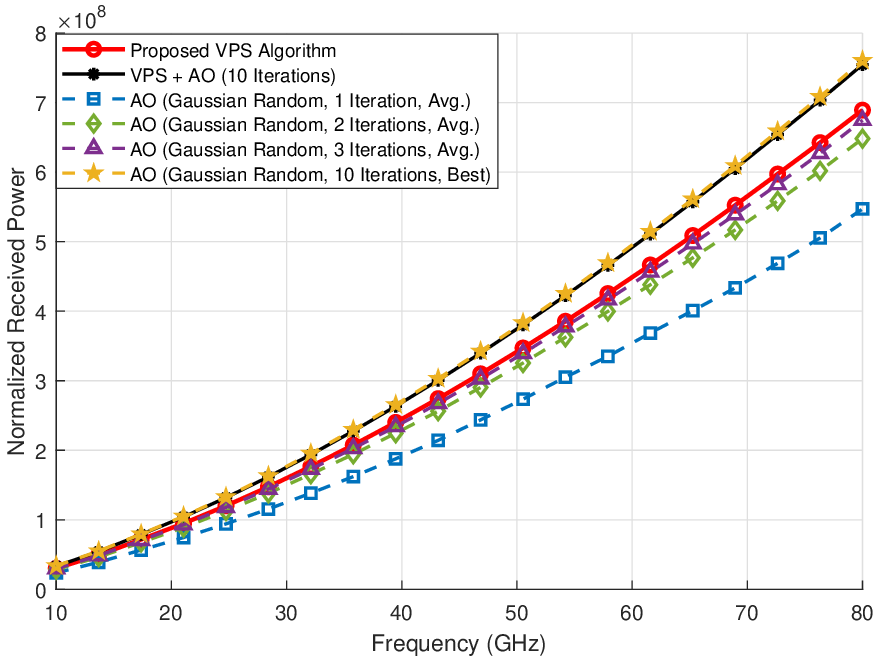}
		\caption{Normalized received power versus frequency with fixed Rayleigh distance.}
		\label{fig_exp4_freq_rayleigh_vs_power_chap5}
	\end{figure}

\begin{figure}[!t]
		\centering
		\includegraphics[width=0.85\columnwidth]{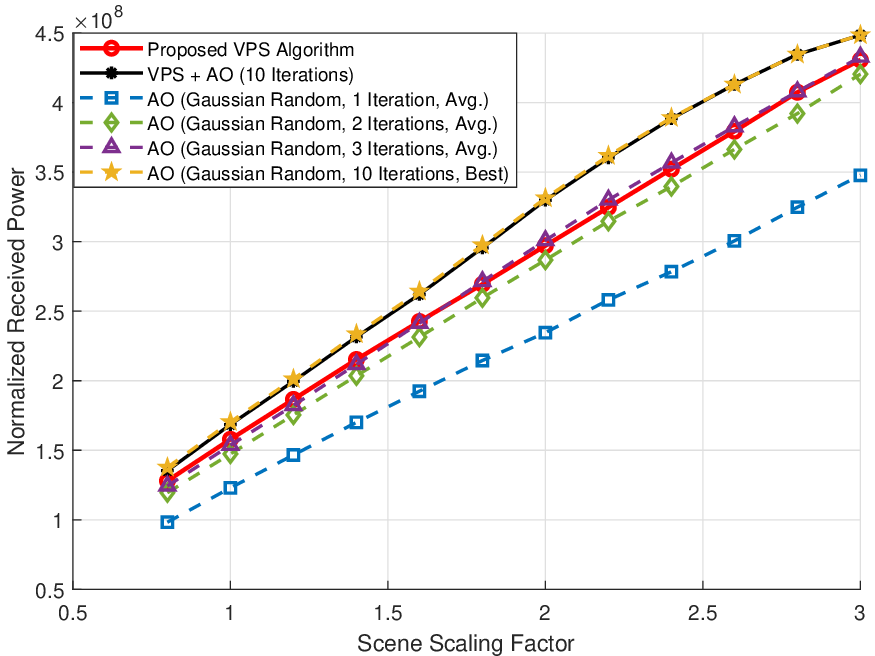}
		\caption{Normalized received power versus the scenario scaling factor.}
		\label{fig_exp5_scale_vs_power_chap5}
	\end{figure}

	In this section, numerical results are provided to validate the performance of our proposed VPS algorithm. We consider an IRS-aided single-user communication system where a multi-antenna BS, an IRS, and a single-antenna UE are all located in a 2D plane. Unless specified otherwise, the baseline scenario adopted for all simulations in Fig.~\ref{fig_ite_HMIMO_chap5} through Fig.~\ref{fig_exp5_scale_vs_power_chap5} is defined as follows: The BS array center is located at the origin $(0, 0)$, with the tangent of its array oriented at a $30^\circ$ clockwise angle relative to the y-axis. The IRS array center is located at $(50\text{m}, 0)$, and its normal vector forms a $60^\circ$ angle with the positive x-axis. The UE is located at $(37.5\text{m}, -12.5\text{m})$. The system carrier frequency is set as $f_c = 30 \text{ GHz}$, which is used for all experiments except where frequency is explicitly varied (i.e., Fig.~\ref{fig_exp1_freq_vs_power_chap5} and Fig.~\ref{fig_exp4_freq_rayleigh_vs_power_chap5}). The BS comprises $N_{BS} = 400$ elements with $\lambda/2$ spacing (physical aperture $L_{BS} \approx 2\text{m}$), and the IRS comprises $M_{IRS} = 2000$ elements with $\lambda/4$ spacing (physical aperture $L_{IRS} \approx 5\text{m}$). All channels are assumed to follow the LoS spherical-wave model. The received power is normalized by $N_{BS}$.
	
	Fig.~\ref{fig_ite_HMIMO_chap5} first illustrates the convergence performance of the conventional AO algorithm under different initialization strategies for the baseline scenario. The non-convex nature of the DNF beamforming problem is evident. We compare the following benchmark initializations: 1) \textbf{Far-Field Focusing:} initializes the AO algorithm by incorrectly assuming planar wave propagation (the far-field model) instead of the correct spherical-wave model; 2) \textbf{Zero Phase:} sets all initial phase shifts at both the BS and IRS to zero radians; 3) \textbf{Uniform Random:} all phase shifts are independently drawn from a uniform distribution $\theta_n \sim U[0, 2\pi)$; and 4) \textbf{Gaussian Random:} the initial phase shifts $\theta_n$ are independently drawn from a Gaussian distribution $\theta_n \sim \mathcal{N}(0, (\pi/4)^2)$ (i.e., mean 0 and standard deviation $\pi/4)$. As shown, these strategies converge to substantially different performance levels. To illustrate the impact of random initialization, two independent trials are plotted for both Gaussian Random and Uniform Random, which confirms that the AO algorithm is highly sensitive to its initial value and prone to converging to local optima. In contrast, the red solid curve corresponds to the AO algorithm initialized by our proposed non-iterative VPS solution. Two key observations can be drawn from this result. First, at iteration 0, the performance of the non-iterative VPS solution (the starting point of the red solid curve) is already significantly higher than the final converged performance of all other benchmark strategies. Second, while all traditional initializations converge to sub-optimal levels, the solution initialized by VPS (the red solid curve) achieves the highest performance upon convergence. This demonstrates that the proposed VPS method is not only an effective non-iterative solution but also serves as an ideal initial value for the iterative algorithm, capable of guiding the AO process to a superior solution and avoiding poor local optima.
	
	After establishing the convergence behavior in Fig.~\ref{fig_ite_HMIMO_chap5}, we further evaluate the performance and robustness of the proposed algorithm in Fig.~\ref{fig_exp1_freq_vs_power_chap5} through Fig.~\ref{fig_exp5_scale_vs_power_chap5}. In these experiments, one specific parameter is varied while all others are kept as in the baseline scenario. We compare the performance of the following schemes: 1) \textbf{VPS:} the proposed non-iterative solution (red solid line with circles); 2) \textbf{VPS + AO:} the converged AO algorithm initialized by the proposed VPS solution (black solid line with asterisks); 3) \textbf{Random AO (Avg.):} the average performance of randomly initialized AO after 1, 2, and 3 iterations (dashed lines with square, diamond, and triangle markers, respectively), based on 100 independent trials; and 4) \textbf{Random AO (Best):} the best performance achieved among the 100 random trials after 10 iterations (yellow dashed line with pentagrams).
	
	From Fig.~\ref{fig_exp1_freq_vs_power_chap5} to Fig.~\ref{fig_exp5_scale_vs_power_chap5}, two consistent observations can be made. First, the non-iterative VPS algorithm (red solid line with circles) is always shown to significantly outperform the average performance of the randomly initialized AO algorithm after 1, 2, or 3 iterations (the blue, green, and purple dashed lines). This demonstrates the high efficiency of the proposed non-iterative solution. Second, the performance of the converged AO algorithm initialized by VPS (black solid line with asterisks) is observed to be almost identical to the best performance achieved among 100 random initializations (yellow dashed line with pentagrams) across all test scenarios. This result strongly validates that the proposed VPS method provides a near-optimal initial value, effectively overcoming the local optimum problem inherent in the conventional AO algorithm without the need for extensive random searches.
	
	Specifically, these five experiments validate the above conclusions under different conditions. Fig.~\ref{fig_exp1_freq_vs_power_chap5} investigates the impact of frequency (10-80 GHz) with a fixed physical aperture, confirming the VPS advantage holds across varying electrical apertures. Fig.~\ref{fig_exp2_bs_angle_vs_power_chap5} and Fig.~\ref{fig_exp3_irs_angle_vs_power_chap5} demonstrate the algorithm's robustness against angular misalignments of $\pm 15^\circ$ at both the BS and IRS. Fig.~\ref{fig_exp4_freq_rayleigh_vs_power_chap5} confirms that these gains are universal and not dependent on a specific degree of ``near-fieldness'' by fixing the Rayleigh distance. Finally, Fig.~\ref{fig_exp5_scale_vs_power_chap5} shows the algorithm's scalability by varying the communication distance (scaling factor 0.8 to 3.0), where the VPS method maintains its performance advantage and optimality as an initial value regardless of the scene scale.

	\section{Conclusion}
	
This paper investigated the near-field beamforming design problem in XL-MIMO systems and proposed a novel holographic paradigm to address the computational complexity and performance bottlenecks of traditional methods. Specifically, as communication frequencies and antenna apertures increase, near-field propagation effects—particularly in the DNF scenario—induce strong coupling among design variables. This renders conventional iterative algorithms, such as AO, highly sensitive to initialization and prone to suboptimal convergence. To address this challenge, we established an analytical framework for XL-MIMO from a continuous-space perspective and introduced the ``holographic methodology,'' which reformulates the high-dimensional, discrete optimization problem into a task of shaping a CWF over the array aperture. Addressing the DNF coupling specifically, we proposed the VPS method. This approach approximates the ideal CWF with an analytically tractable spherical-wave and provides a rigorous geometric-optical proof that the optimal VPS location can be determined non-iteratively via the ``opposing triangles'' method, thereby efficiently decoupling the beamforming design. Simulation results in an IRS-assisted system validate that the proposed non-iterative VPS algorithm achieves performance comparable to, or even exceeding, that of converged conventional AO algorithms, while significantly reducing computational complexity and eliminating iterative uncertainty. This work offers a new theoretical framework and a practical solution for XL-MIMO near-field communications, providing vital insights for the design of continuous-aperture arrays in future 6G systems.
	\ifCLASSOPTIONcaptionsoff
	\newpage
	\fi

	%
	%
	\bibliographystyle{IEEEtran}
	\bibliography{HMIMO}  

\end{document}